\documentclass[a4paper,fleqn,usenatbib]{pazhbabel}
\usepackage{fullpage}
\usepackage{graphicx}
\usepackage{morefloats}
\usepackage{lscape}
\usepackage{fancyhdr}
\usepackage{amsmath}
\usepackage[koi8-r]{inputenc}
\usepackage[russian]{babel}
\usepackage[T2A]{fontenc}
\usepackage{caption}

\begin{document}

\journalinfo{2015}{10}{0}{1}[0]

\title{Peculiarities of the Accretion Flow in the System HL CMa}
\author{\bf %\hspace{-1.3cm} \copyright
A.~Semena \email{san@iki.rssi.ru}\address{1}, M.~Revnivtsev\address{1}, D.~Buckley\address{2}, A. Lutovinov\address{1},\\  H.~Breytenbach\address{2}
\addresstext{1}{Space Research Institute, Russian Academy of Sciences,Profsoyuznaya ul. 84/32, Moscow, 117997 Russia}
\addresstext{2}{South African Astronomical Observatory, P.O. Box 9, Observatory, Cape Town 7935, South Africa} %, Южная Африка}
}
\shortauthor{}
\shorttitle{}
\submitted{\today}
\thispagestyle{empty}

\begin{abstract}

The properties of the aperiodic brightness variability for the dwarf nova HL CMa are considered.
The variability of the system HL CMa is shown to be suppressed at frequencies above $7\times10^{-3}$Hz.
Different variability suppression mechanisms related to the radiation reprocessing time, partial disk evaporation, and characteristic variability formation time are proposed. 
It has been found that the variability suppression frequency does not change when the system passes from the quiescent state to the outburst one, suggesting that the accretion flow geometry is invariable. 
It is concluded from the optical and X-ray luminosities of the system that the boundary layer on the white dwarf surface is optically thick in both quiescent and outburst states. 
The latter implies that the optically thick part of the accretion flow (disk) reaches the white dwarf surface. 
The accretion rate in the system, the flow geometry and temperature have been estimated from the variability power spectra and spectral characteristics in a wide energy range, from the optical to X-ray.

\end{abstract}

\label{firstpage}

\section{Introduction}

Cataclysmic variables are close binary systems that consist of a white dwarf and a less massive Roche-lobe-filling companion star.
These systems have their name from the observed rapid changes in their optical brightness by several magnitudes \citep[for a review, see][]{warner03}. 
One of the characteristic features of these systems is a significant
brightness variability amplitude compared to single
stars in the frequency range from $ 10^{-4}$ to several Hz \citep[][]{linnell50}. 
In addition to cataclysmic variables, close binary systems in which neutron stars and black holes are accretors also show a large variability amplitude. 
The common properties of the variability in these systems suggest that the variability of their luminosity is formed by accretion disks.
	
Some of the known cataclysmic variables exhibit regular aperiodic outbursts in which their optical brightness rises by 2-5 magnitudes. 
The systems in which such outbursts are observed belong to a subclass of cataclysmic variables called dwarf novae.
The duration of outbursts in these objects is typically several days, while the interval between them can be from several weeks to several months \citep[][]{warner03}.

It is believed that the accretion flow can be inhomogeneous -- part of the disk in the system can be evaporated and be optically thin with a high viscosity \citep[see, e.g.][]{king97}.
The properties of the flow in the accretion disk are  reflected in the power spectrum of its luminosity.
Dwarf novae are particularly interesting in studying the variability properties of the accretion flow, because an accretion disk in various physical states can be observed in the same system. 
The change in the shape of the luminosity power spectrum for the system when it passes from the quiescent state into outburst will allow the changes in disk geometry to be determined.

In this paper, based on observational data for the dwarf nova HL CMa, we analyze the variability of its optical luminosity in the quiescent and outburst states. We estimate the accretion rate in the system as well as the accretion flow geometry and temperature from the variability power spectra and spectral characteristics in a wide energy range (from the optical to X-ray).

\section{ACCRETION FLOW IN DWARF NOVAE}
	
Accretion disk instability is believed to be responsible for the outbursts in dwarf novae \citep[for a review, see, e.g.,][]{lasota01}. 
The accretion disk instability model attributes the outbursts of dwarf novae to a sharp increase of the accretion rate in the disk.
The simplest model in which the outburst behavior of binary systems is reproduced is the so-called one-zone accretion disk model.
	
Disk instability is attributed to hydrogen ionization and a subsequent change in the vertical structure of the disk when the necessary critical density is reached in some part of it \citep{meyer81}. 
In the region where the critical density is reached, the accretion rate increases, causing the matter to be heated further. 
In this case, a heating wave propagates away from this region over the disk, which causes the accretion rate to increase in other parts of the disk.
Once the disk has lost part of its mass accumulated during the quiescent phase, the matter in it reaches another critical density (lower than that at which the transition to the hot state occurred) and returns to the cold state with a low viscosity, while the accretion rate returns to its pre-outburst level \citep{mineshige83}. 

Instability can begin in different parts of the disk.
Depending on this, the outburst profile will change in different wavelength ranges. 
Numerical simulations within the model described above yielded light curves similar to those of dwarf novae, including two characteristic outburst profiles -- for the cases of instability onset in the outer and inner parts of the disk \citep[see, e.g.][]{smak84, cannizzo93}.
	
The one-zone model allows the outbursts of dwarf novae to be described qualitatively, but a number of effects observed during outbursts remain unexplained. 
In particular, a delay of $0.5 --1.5$ days in the rise in ultraviolet luminosity as compared to the optical is observed at the onset of outbursts in most dwarf novae \citep[see, e.g.][]{verbunt87_1, polidan87_1}.

It is believed that the ultraviolet emission in dwarf novae must originate from the hottest inner part of the optically thick accretion disk as well as from the white dwarf surface and the boundary layer \citep{diaz96,popham95,piro04}. 
The delay of the ultraviolet outburst relative to the optical, points to a delay of the rise in the accretion rate in the inner parts of the disk relative to the outer parts. 
Such a delay is expected in the standard one-zone disk instability model due to the finite propagation time of the heating wave from the outer parts of the disk into its inner regions. 
However, the characteristic propagation time of the heating wave turns out to be less than 0.5 day. 
\citet{smak98} attempted to explain the observed delays within the context of the one-zone model, but the boundary and initial conditions in his modeling were unrealistic.

To solve the problem of a prolonged ultraviolet outburst delay, a model in which there is no inner part of the optically thick cold disk was proposed \citep[see, e.g.][]{livio92}. 
In this model, the delay is determined not by the propagation time of the heating wave over the disk but by the travel time of the inner disk edge to the white dwarf surface. 
The absence of an inner part of the cold optically thick disk can be explained by the evaporation of matter from the disk into a hot optically thin corona \citep[see, e.g.,][]{meyer94,king97,dullemond99}. 
In this case, the accretion onto the white dwarf surface will occur in a hot optically thin accretion flow.
	
\begin{figure}
\includegraphics[width=1\columnwidth]{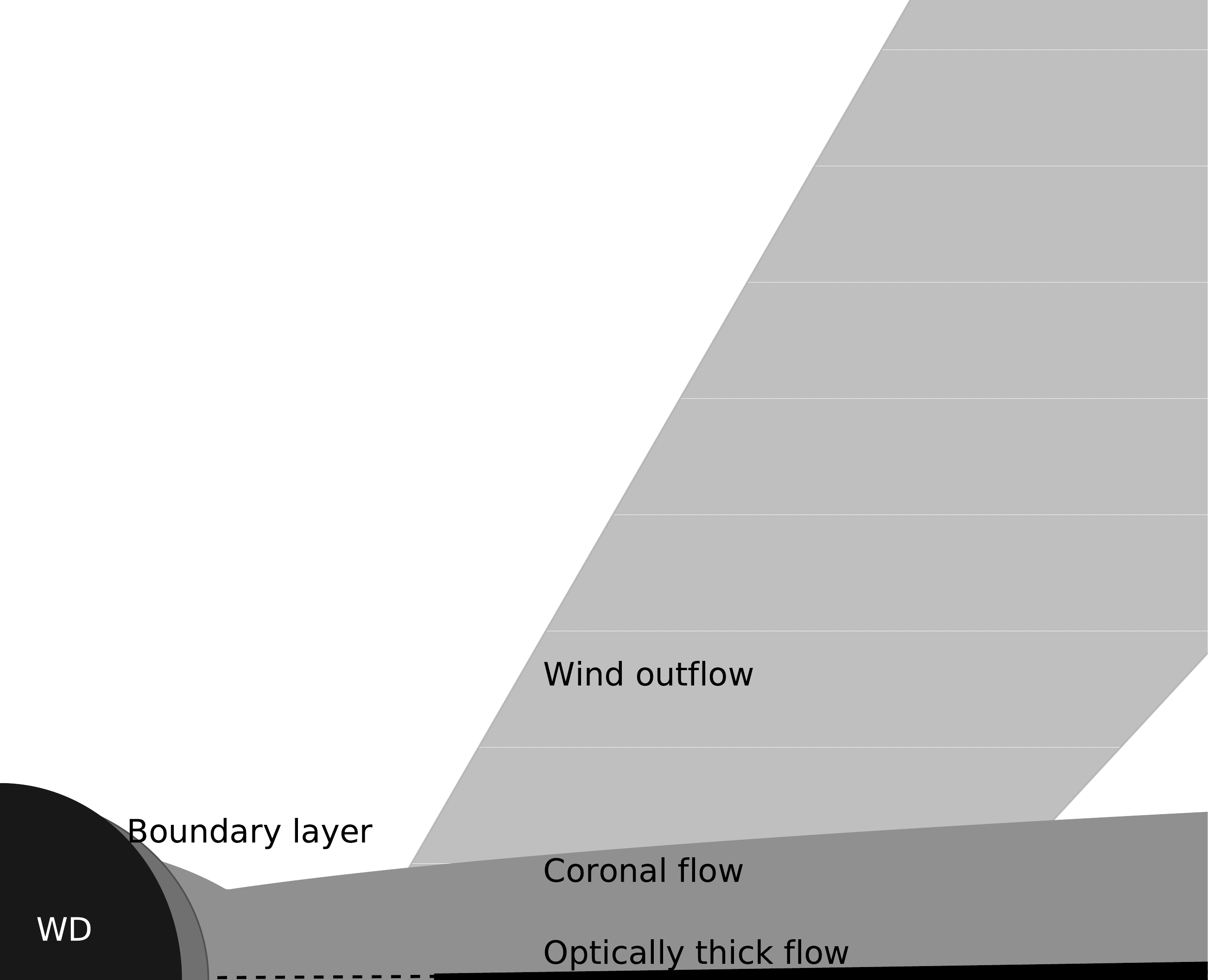}
\caption{
Schematic view of the accretion flow near the white dwarf in nonmagnetic cataclysmic variables. In a dwarf nova in its quiescent state, it is expected that the inner part of the optically thick disk has been evaporated and the accretion goes in an optically thin corona forming above the disk (see the text). 
A wind is also formed in the system in its outburst state. If the evaporation efficiency is lower than the accretion rate in the optically thick disk, then it reaches the white dwarf (WD) surface.}
\label{fig:b_scheme}
\end{figure}

Thus, based on the model described above, the accretion flow around the white dwarf in a typical dwarf nova can be represented as a combination of an optically thick, geometrically thin, cold disk, the inner edge of which may not reach the white dwarf surface; and a hotter optically thin flow forming above the disk and extending to the surface of the WD (fig.\,\ref{fig:b_scheme}). 
It should be noted that the inner radius of the optically thick accretion disk will be determined by the evaporation efficiency with respect to the accretion rate in it. 
The disk can be almost completely evaporated (in the case of a high evaporation efficiency) or can reach the white dwarf surface.
In the latter case, only part of the accretion flow will go in the optically thin flow.

%==============================================================	
%==============================================================	

The spectra of some dwarf novae and the light curves of eclipsing systems in various energy ranges  also point to the absence of a part of the optically thick accretion disk inside a region with a size of several white dwarf radii \citep[see e.g.,][]{schoembs81, berriman85, polidan87_2, wood92, hoard97, dous97, gaensicke99, biro2000, mcgowan04, hartley05, long05, belle05, linnell07, puebla07, revnivtsev10}.
	
One of the goals of this paper is to determine the sizes of the evaporated part of the disk and, accordingly, the evaporation efficiency in the system HL CMa.
	
\section{APERIODIC OPTICAL LUMINOSITY VARIABILITY}
		
The properties of the flow in the accretion disk can be used to estimate the inner radius of the optically thick disk by analyzing the temporal characteristics of the brightness variability in a binary system. 
The brightness variability of close binary systems in a wide frequency range is known to be formed in their accretion disk (see, e.g., the model of propagating fluctuations proposed by \citep[see, e.g., proposed by][model of propagating fluctuations (hereafter L97)]{Lyubarski97}. 
In this model, the luminous variability is generated by a variable accretion rate in the disk which, in turn, is formed by stochastic fluctuations in viscosity \citep{balbus91, brandeburg95, hirose06}. 
As the matter moves in the accretion disk, independent perturbations are superimposed on the accretion rate, giving rise to a flux variability power spectrum at the inner disk edge which has a power law form $P \approx f^{\alpha}$, where $\alpha = -1...-2$.

The accretion rate perturbations produced by the disk have certain time scales. A number of observational manifestations, such as the change of the power spectra with increasing accretion rate in magnetic systems and the form of the correlation function of the optical and X-ray emissions from intermediate polars, suggest that the power spectrum of the accretion rate variability is formed on the Keplerian time scales of the optically thick accretion disk \citep[see, e.g.][]{revnivtsev09, revnivtsev10, revnivtsev11}). 

Given finite disk sizes, variability in the rate of the flow that reaches the white dwarf surface must be suppressed at frequencies above the characteristic formation frequency of perturbations at the inner edge of the optically thick disk. 
Thus, it follows from the theory of propagating fluctuations that the frequency at which the power spectrum passes from a power law with a slope $\alpha \simeq -1$ to a new law with a larger slope corresponds to the Keplerian frequency at the inner edge of the optically thick disk.

The X-ray light curves are best suited to determining the properties of the accretion flow, because the bulk of the energy release in dwarf novae in this wavelength range occurs near the white dwarf surface, where the flow variability power spectrum has already been formed. 
However, because of the low photon flux, the X-ray and ultraviolet observations of dwarf novae aimed at analyzing the power spectra generally require tens or hundreds of kiloseconds and, therefore, are not always possible.

\citet{revnivtsev11} showed the optical brightness variability of intermediate polars to also be determined by the energy release near the white dwarf surface. 
The optical luminosity variability of the disk itself turns out to be lower than the variability of the optical component that results from the reprocessing of radiation from the inner part of the accretion flow intercepted by the white dwarf and accretion disk
surfaces.

A significant difference in the radiation mechanisms of dwarf novae in the quiescent and outburst states should be noted. 
In the quiescent state, an optically thin boundary layer with a temperature of several keV emitting predominantly in the X-ray energy band, in which the matter cooling time is less than 10 s, is observed in many systems \citep{pringle79}. 
The X-ray reprocessing time into optical emission also turns out to be short \citep{cominsky87}.

In the case of a high accretion rate, the boundary layer becomes optically thick. The cooling time scale for the optically thick boundary layer is less than 50 sec \citep[see Eq. (45) from][]{piro04}. 
The temperature of the optically thick boundary layer is $(2-5) \times 10^5$ К, and the bulk of the flux emitted by it is accounted for by the hard ultraviolet and soft X-ray parts of the spectrum.
Therefore, the characteristic reprocessing time of the radiation from the system's central regions may turn out to be much greater than that in the case of an optically thin boundary layer \citep[for the reprocessing of hard ultraviolet radiation in accretion disks, see][]{suleimanov99}. 
This can lead to changes of the variability power spectrum in the optical range with respect to the variability power spectrum in the hard ultraviolet and X-ray parts of the spectrum.

However, it is worth noting that the power spectrum of the optical brightness for the system SS Cyg in its bright state shows a high variability amplitudes up to frequencies $\sim 0.1$ Hz \citep{revnivtsev12} despite the fact that the boundary layer in this system in its bright state is optically thick.

The observed brightness variability power spectrum for dwarf novae in a wide range of Fourier frequencies has the form of several power laws passing into one another (also occasionally interpreted as a combination of broad Lorentzians with certain frequencies) and characteristic peaks corresponding to different periods in the systems \citep[see, e.g.,][]{scaringi12}. 
We assume that the break (the frequency at which the frequency dependence of the power passes from a power law with a slope -1 to a power law with a larger slope) in the high-frequency part of the spectrum (at frequencies above $10^{-3}$ Hz) is associated with the inner edge of the optically thick accretion disk. 

\section{OBSERVATIONS}	
	
Hundreds of dwarf novae are known at present \citep{ritter14}, but their irregular behavior makes it difficult to carry out the observations aimed at obtaining the spectra and time series in the quiescent and bright states. 
To investigate the accretion disk evolution, we chose the star HL Canis Majoris (HL CMa), one of the brightest dwarf novae, discovered in 1981 by the Einstein X-ray observatory. 
This cataclysmic variable shows regular outbursts every 15?20 days.

The goal of the observations of the cataclysmic variable HL CMa is to determine the properties of its variability in the frequency range from several $10^{-3}$Hz to about $0.5$ Hz in the quiescent and outburst states. 
These frequencies correspond to the Keplerian times of the accretion disk $T = 2 \pi \sqrt{R^{3}/(G M_{wd})}$ or
a standard cataclysmic binary system, where $M_{wd}$ -- is the white dwarf mass, $R$ -- are possible radii of the inner edge of the optically thick disk in cataclysmic variables, $ R_{wd} < R < 0.8R_{rl}$ and $R_{rl}$ -- is the Roche lobe radius. For HL CMa, based on $M_{wd} = M_\odot$, $q = 0.5$ \citep{hutchings81} and $T_{orb} = 0.21678$ day \citep{sheets05}, $R_{rl} \approx 5.3 \times 10^{10}$cm.  
Thus, there are orbits with a radius of 0.4 - 0.25 of the white dwarf radius in the investigated range of Keplerian frequencies.

The observations were carried out with the 1.9-m optical telescope of the South African Astronomical Observatory (SAAO). 
An ANDOR iXon 888 DU CCD array (http://www.andor.com/scientific-cameras/ixon-emccd-camera-series/ixon3-888) \cite{coppejans13} is used in this telescope as a detector. 
The CCD image corresponds to an angular size in the sky of ${1.29}' \times {1.29}'$. 
The instrument makes it possible to carry out observations at frequencies up to 300 Hz.
	
The visual (V band) magnitude of HL CMa is 14.
During the observations, the sky field was chosen in such a way that the instrument's field of view included several constant stars. 
Apart from HL CMa, one or two stars with magnitudes of $\sim 14$, can fall within the instrument's field of view; the remaining stars in the telescope's field of view are dimmer.

\begin{figure}
\includegraphics[trim=0.25cm 0.6cm 0.2cm 0.2cm, clip, width=1.\columnwidth]{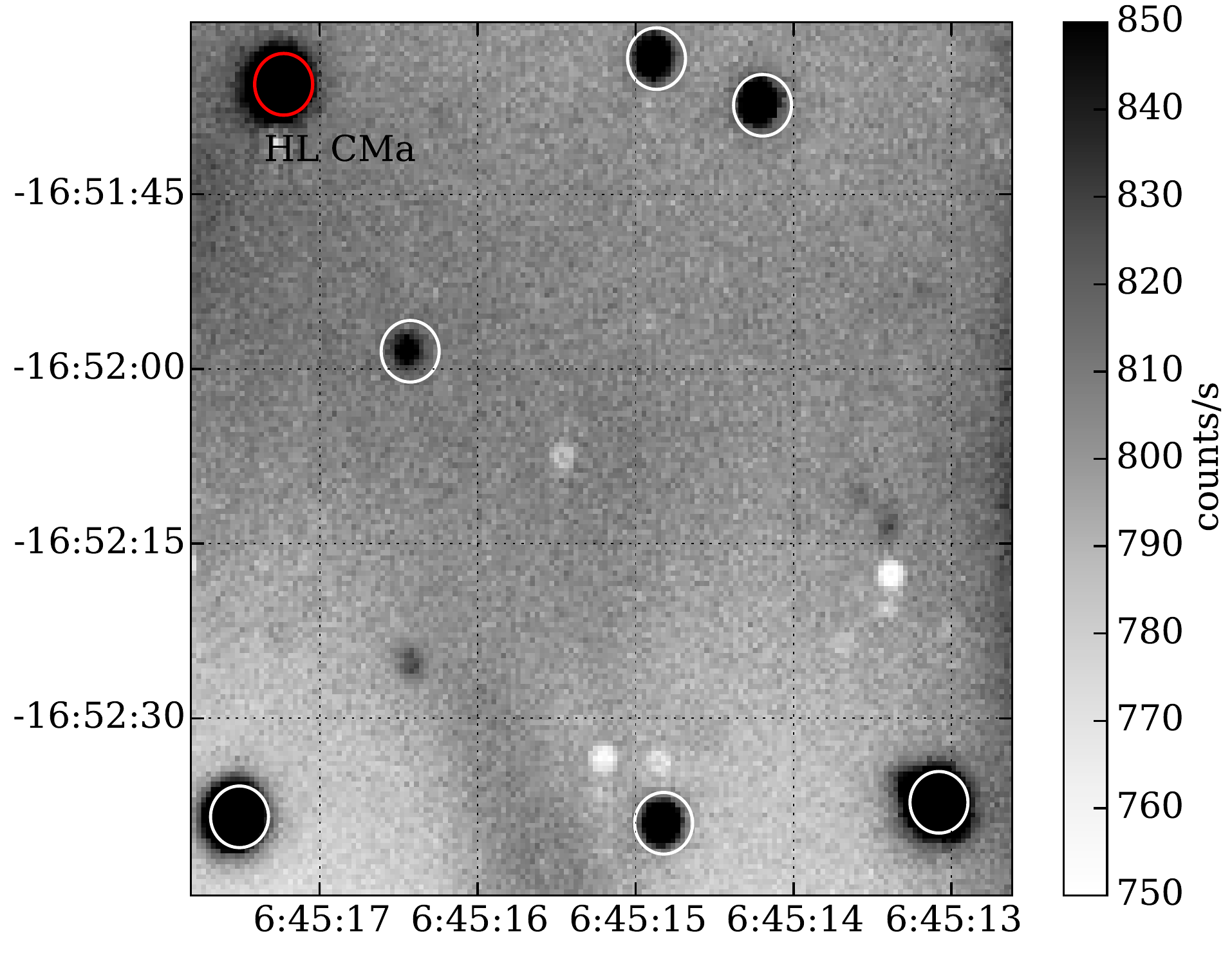}
\caption{Image of the sky field observed with the 1.9-m SAAO telescope to obtain the light curves of HL CMa.}
\label{img:star_fk5}
\end{figure}

The cataclysmic variable HL CMa is 8 arcmin away from Sirius, the brightest star in the sky. 
Such a location of HL CMa makes it difficult to obtain its optical spectra and light curves, because the emission from Sirius is reflected from the telescope's structure, which gives rise to a complex nonuniform background in the instrument's field of view (see Fig.\ref{img:star_fk5}).

%=============================================================	
%==============================================================

\section{LIGHT CURVE CONSTRUCTION METHOD} 	

The result of the CCD operation is a set of images for the selected sky region. 
The number of counts (brightness) of each pixel in the image being obtained is proportional to the number of photons falling into the pixel. 
An example of such an image made from 12 frames, each with an exposure time of 1 s, is shown in Fig\,\ref{img:star_fk5}. 
Because of the photon scattering in the Earth?s atmosphere, the stellar image at ground-based telescopes has a significant angular size \citep[for astronomical seeing, see, e.g.,][]{coulman85}. 
The radius of the aperture including half of the photons from the star during our observations was $\approx 2''$, which is slightly poorer than the median value for the SAAO, $\sim {1.32}''$ \citep{catala13}. 
In view of the weather conditions, the CCD resolution was set lower than the maximum available $1024 \times 1024$, which allowed the dead time and the data volume to be reduced (we used a resolution from $102 \times 102$ to $170 \times 170$, depending on the weather conditions).

When present-day CCD cameras are used, the light curves of observed stars are measured from the number of counts on the CCD image in the aperture containing the star being studied. The aperture sizes should be chosen in such a way that the number of counts from the star in it is maximal at the minimal background contribution. One should also take into account the fact that the number of background counts in the aperture increases linearly with aperture area, $N_{bkg} \propto S_{app}$, while the growth rate of the number of counts from the star decreases with aperture size,$N_{star} \propto S_{app} - \epsilon S_{app}^2$ (the star's profile can be fit by a Gaussian in a small neighborhood of its image center). 
Since all of the photons falling into the aperture are independent, the standard deviation of the luminosity estimate obtained (proportional to the number of counts) is proportional to the square root of the total number of counts in the aperture. 
The best quality of the light curve is achieved at the largest ratio of the signal from the star in the aperture to the standard deviation of the total signal in the aperture $max(N_{star}/\sqrt{N_{star} + N_{bkg}})$. 
The distribution of the background counts can differ from the Poissonian one, and the presented equations will not yield the correct result. Therefore, to find the best aperture size, we measured the standard deviation of the star $\sigma_{star}$ and the background $\sigma_{bkg}$ for various aperture sizes and chose the aperture that satisfied the requirement $max (\sigma_{star} / \sigma_{bkg})$. 
In most cases, the optimal aperture size in our observations was ${3} - {3.7}''$. 
For simplicity, when constructing the light curves for all stars, we chose the optimal aperture size according to the algorithm described above (see the table).

The CCD background model (the background was approximated by inclined planes near the star outside the aperture) was subtracted from the image inside the aperture.

Apart from the background, which generally has no intrinsic variability, the images being obtained can be affected by weather conditions (for example, the changes in atmospheric transparency) and technical peculiarities of the instrument with which the observations are carried out (for example, a change in CCD temperature). 
These effects will lead to variations of the emission being recorded from the stars, but they are identical for all stars on the image.
Therefore, to reduce the amplitude of the variability produced by these effects, it will suffice to use the method of differential photometry. 
In this method, the light curve of the star being investigated is considered to be the ratio of the total number of its counts to the number of counts from the comparison star. 
In the case of HL CMa, there were no bright constant stars in the telescope's CCD visibility region (all stars had a brightness comparable to or lower than that of HL CMa) and the combined flux from several stars that fell within the field of view of the telescope's CCD was used as a comparison star.
		
%====================================================================
\begin{figure}
\includegraphics[trim=1.2cm 2.0cm 0.4cm 2.0cm, clip, width=1.\columnwidth]{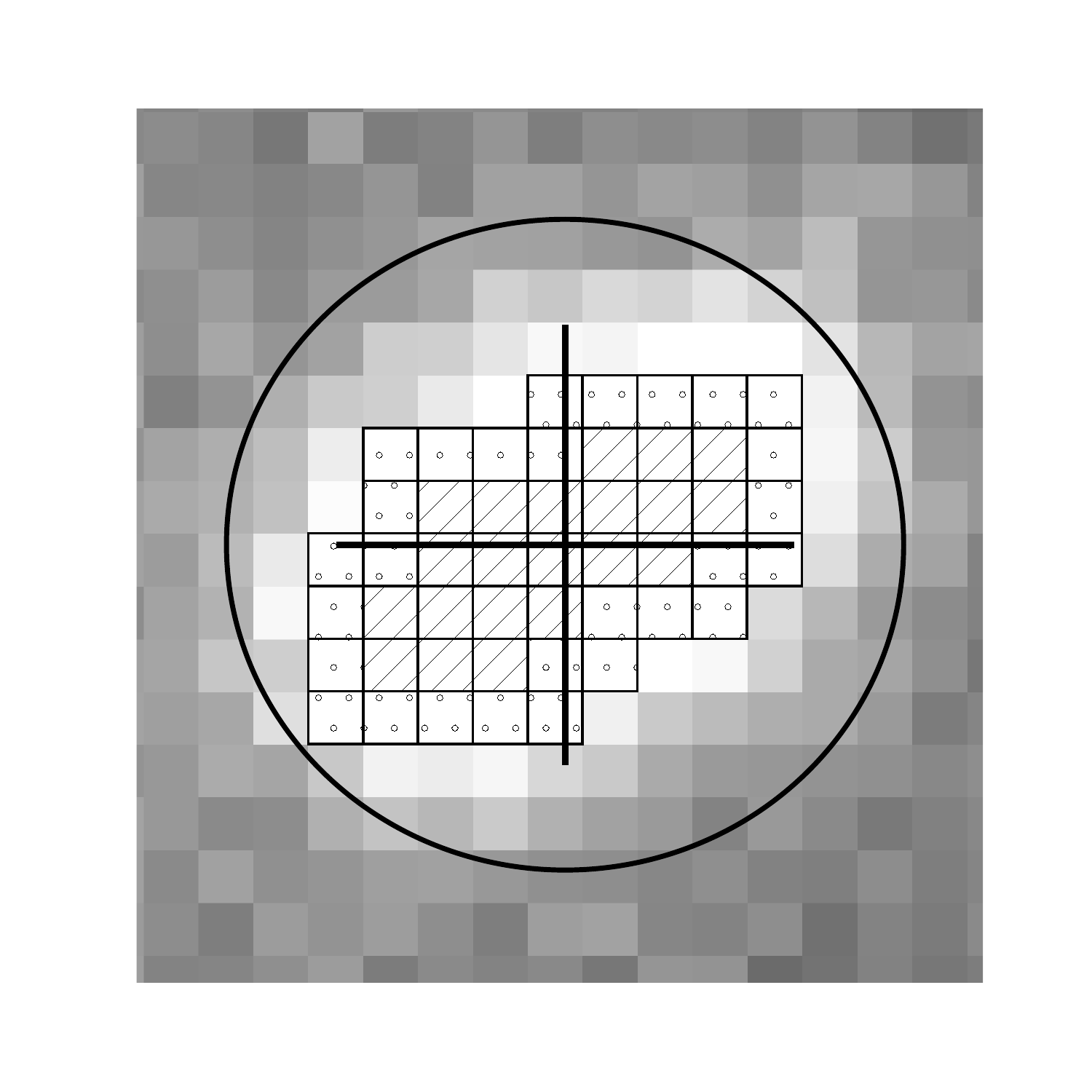}
\caption{Example of the operation of the algorithm for determining the flux from a star with an asymmetric profile. 
The hatched squares represent the pixels in which the photons from the star will be counted. In the next step, the algorithm will choose the brightest pixel at the boundary with the hatched region (the boundary is represented by the region with dots) and will add it to the group of pixels in which the signal from the star will be summed. 
The expansion is limited by the aperture whose size is larger than the optimal one (indicated by the circle). 
The algorithm ends its operation once N pixels whose area is equal to the optimal aperture area have been chosen. 
One of the images of a nonvariable star in the January 16, 2015 observation is shown on the frame.}
\label{fig:expand_algorithm}
\end{figure}		
%====================================================================

Apart from the constant component corresponding to Poissonian noise from a finite number of recorded photons, the power spectrum of the constructed differential light curve contains the variability associated with the delays and scatterings of photons from the star in the Earth?s atmosphere.
This atmospheric effect leads to a random change in the brightness and shape of stars in ground-based observations. 
As a rule, the variability associated with the atmospheric effect has the form of a plateau with a turnover at the characteristic frequency determined by atmospheric properties and can be fitted by a simple analytic function $P(f) \propto (1 + (f/f_{atm})^2)^{-0.3}$, where $f_{atm}$ -- is the atmospheric variability suppression
frequency \citep[see, e.g.,][]{dravins98, osborn15}.

In some observations, it was noted that the stellar profile became asymmetric in CCD coordinates and acquired an elliptical shape that transformed back to a circular shape with time (on a time scale of several seconds). 
Such behavior gave rise to additional variability that was not suppressed in the differential light curve. This variability can be suppressed quite efficiently if the shape of the stellar profile in CCD coordinates is somehow taken into account.
		
We used the following method to solve this problem. 
We chose an aperture larger than the optimal one but smaller than the background shape estimation region to obtain the counts from the star. 
In the chosen aperture, we selected the brightest pixels obtained after the subtraction of the background model. 
Such pixels were chosen by the method of expansion to bright pixels starting from the aperture center, i.e., we first chose the pixel at the aperture center, then the brightest of the eight pixels adjacent to it, then the brightest one adjacent to these two pixels, and so on (Fig.\ref{fig:expand_algorithm}). 
Thereafter, the counts in N brightest pixels were summed, while the number of pixels was chosen in such a way that their area was equal to the area of the optimal circular aperture. 
With such a choice of pixels, the shape of the region in which the counts from the star are taken into account can be highly asymmetric, following the shape of the stellar image.
		
%===========================================================================			
\begin{figure}
\includegraphics[width=1.1\columnwidth, trim=0.7cm 1cm 0 1cm]{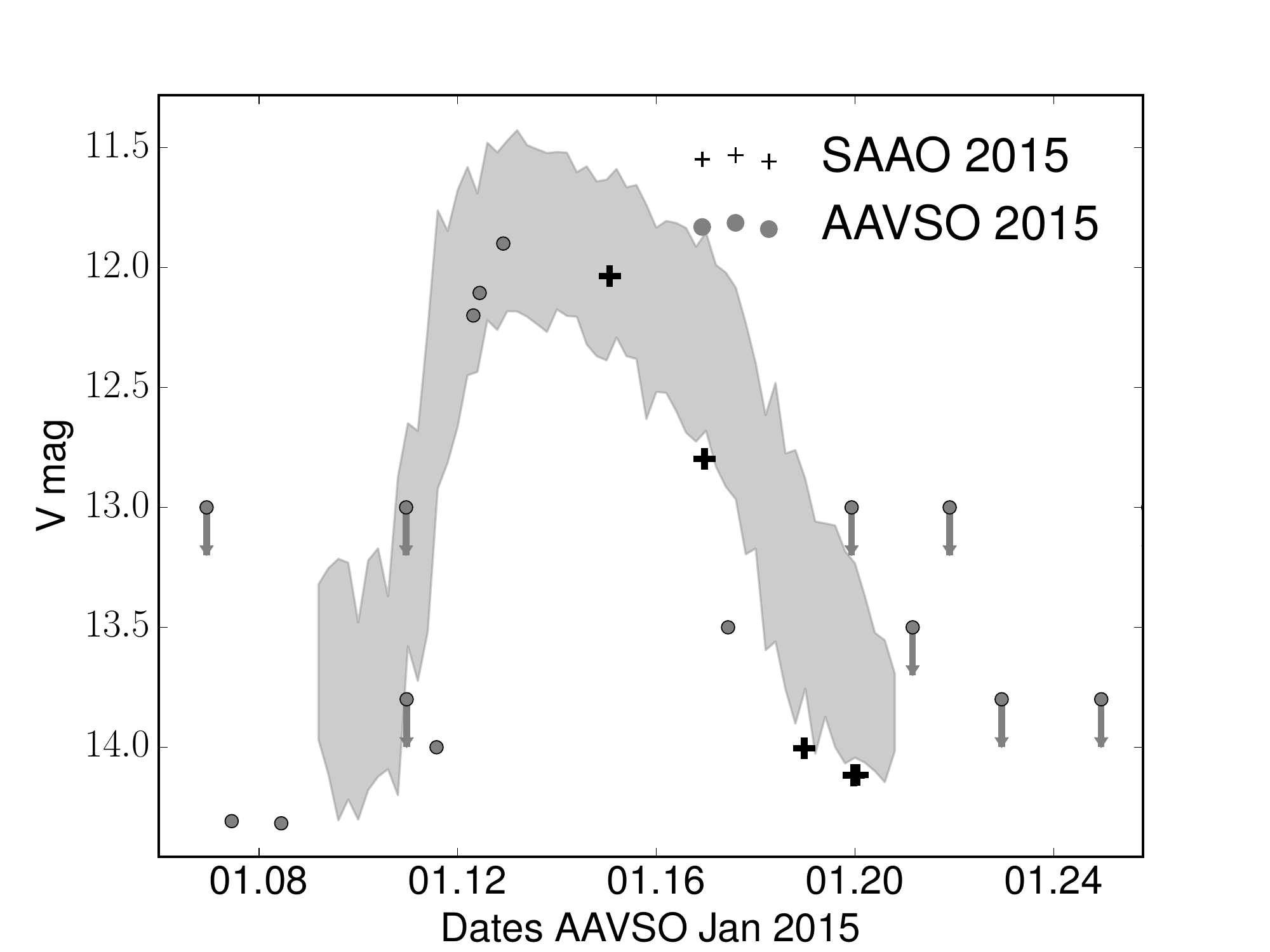}	
\caption{Visual magnitude of HL CMa during its observations in January 2015. The graph shows the magnitudes measured by AAVSO observers (https://www.aavso.org, gray dots) and those measured during the observations at the SAAO telescope (black crosses). 
The gray shaded region represents the characteristic profile of a short outburst for HL CMa obtained by the phase tracking method using AAVSO observations in 2012--2015.}
\label{fig:obs_graph}
\end{figure}
%==========================================================================

The night-sky brightness during our observations was $\approx 18.5 m/arcsec^2$, the mean brightness of illumination from Sirius was $\approx 20.2 m/arcsec^2$. 
The illumination from Sirius has no significant intrinsic variability and leads only to an increase in the amplitude of white noise due to the background brightness rise. 
The background variability amplitude in a circular aperture for a star of magnitude 14 does not exceed 0.3\%. 
Thus, the background does not contribute significantly to the variability amplitude of the stars in the entire range of frequencies of interest to us.

%==========================================================================
\begin{table*}
\centering
\caption{Dates, durations, mean magnitudes, and aperture sizes for the observations performed at the 1.9-m SAAO telescope and used to analyze the variability of HL CMa}
\begin{tabular}{ccccc}
\hline
 Date & Beginning of observations, & Duration, & Magnitude, &  Aperture,\\
      &        UTC        &    sec        &         V         &  arcsec  \\

 \hline
 15.01.15 & 00:57 & 3600 & $12.0$ & 3.7 \\
 16.01.15 & 22:44 & 5400 & $12.8$ & 4.3 \\
 18.01.15 & 22:53 & 7200 & $14.0$ & 3.7 \\
\hline
\label{tbl:obs_log}
\end{tabular}
\end{table*}

In summary, we note that at the beginning of our observations (January 13, 2015), HL CMa was apparently in its outburst state for about two days (Fig.\ref{fig:obs_graph}). 
HL CMa reached a V magnitude of 12 during these observations compared to a magnitude of $\sim 14$ in the observations performed $\sim 6$ days after the outburst onset (see the table).

%======================================================================	
\begin{figure*}
\includegraphics[width=\textwidth, trim=2cm 0 0 0]{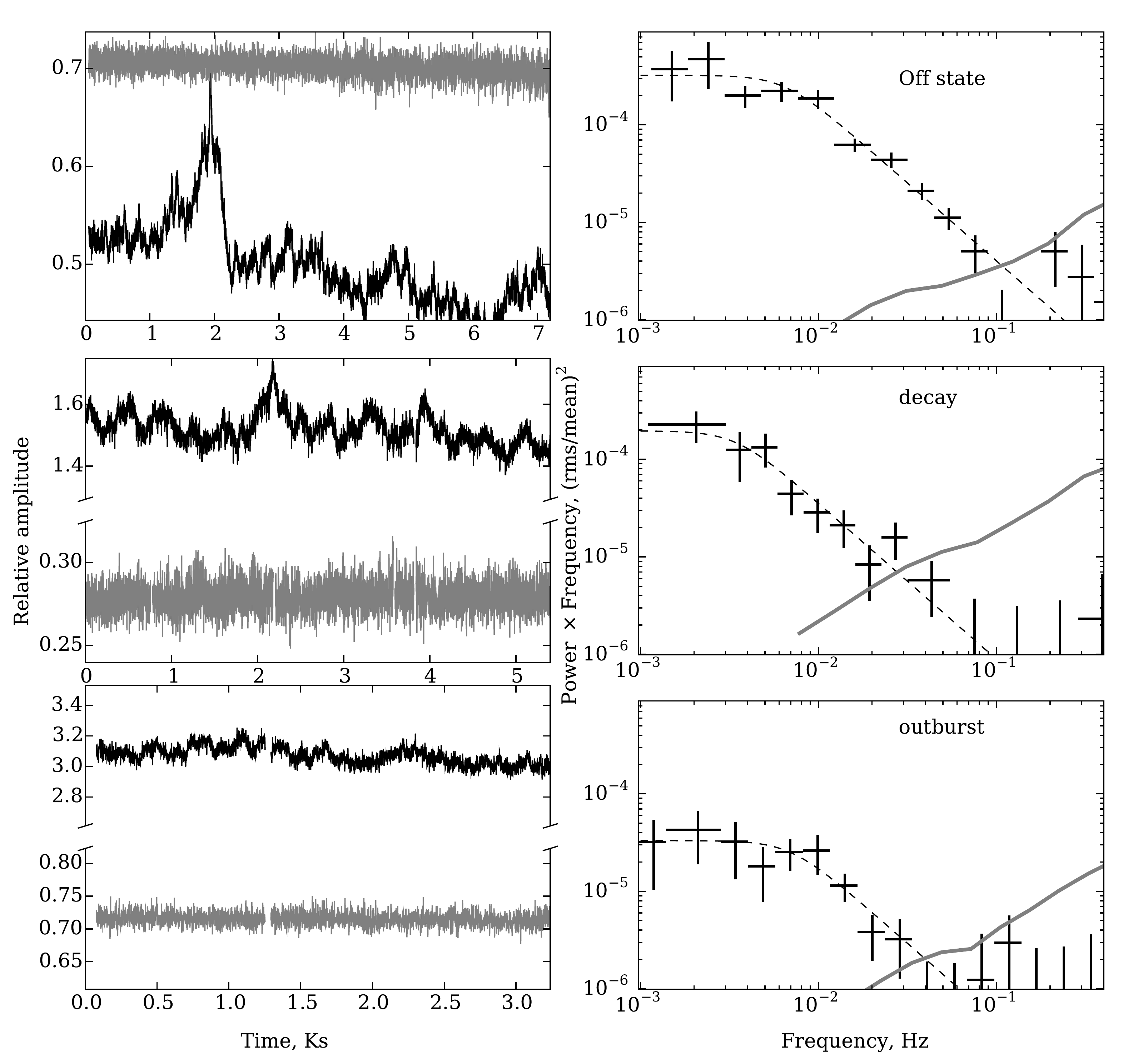}
\caption{The light curves (left) and power spectra (right) of HL CMa in the off (top), decay (middle), and outburst (lower panel) states are shown in pairs. 
The break in the power spectrum corresponding to the transition from a power law with a slope of -1 to a power law with a slope of ?2.6 is located in the frequency range $(0.3 - 1) \times 10^{-2}$ Hz. 
The break frequency six days after the outburst onset (upper panel) is compatible with the break frequency in the bright state $(7.0 \pm 2.0) \times 10^{-3}$ Hz, while the break in the power spectrum in the decaying part of the light curve (four days after the outburst onset) is slightly shifted toward low frequencies $(3.5 \pm ^{1.2}_{1.0}) \times 10^{-3}$ Hz.}
\label{fig:hl_cma_ps}
\end{figure*}
%==================================================================

\section{RESULTS}

The light curves and power spectra of the brightness variability for HL CMa two, four, and six days after the outburst onset are presented in Fig.\,\ref{fig:hl_cma_ps}. 
The transition from a power law with a slope $\alpha \simeq-1$,to
a power law with a slope $\alpha \simeq -2.6$ is clearly seen in the power spectra in the outburst and quiescent states. 
To determine the break frequency, the power spectra were fit by a model including a power law with a break, an atmospheric noise power model, and a constant to describe the contribution of Poissonian noise. 
To estimate the shape of the model function describing the atmospheric variability, the power spectra of the differential light curves for HL CMa and the constant comparison star were fit simultaneously by the maximum likelihood function:

\begin{equation}
\begin{split}
 LKL_{HL CMa}(n, f_{break}, n_{atm}, f_{atm}, p_{HL CMa}) +
 \nonumber
 \\ LKL_{cs}(n_{atm}, f_{atm}, p_{cs}),
 \nonumber
\end{split}
\end{equation}	
where $LKL_{HL CMa} = \sum_i \log{m(f_i)} + p_i/m(f_i)$ -- is the
likelihood function for the power spectrum of HL CMa.
The model $m$ is
\begin{equation}
\begin{split}
 m(f) = n f^{-1} (1 + (f/f_{break})^4)^{-0.4} + \\
 \nonumber
 n_{atm} (1 + (f/f_{atm})^{2})^{-0.3} + p_{HL CMa},
 \nonumber
\end{split}
\end{equation}
where $f_i$, $p_i$ -- are the frequency and power from
the variability power spectrum of the differential
light curve for HL CMa. The expression $LKL_{cs} = \sum_j \log{m(f_j)} + p_j/m(f_j)$ is the likelihood function of the comparison star. The corresponding model is $$m(f) = n_{atm} (1 + (f/f_{atm})^{2})^{-0.3} + p_{cs},$$ where $f_j$, $p_j$ -- are the frequency and power from the variability power spectrum of the differential light curve for the comparison star.
	
The shape of the model function describing the variability of HL CMa, $nf^{-1}(1 + (f/f_{break})^4)^{-0.4}$, was chosen by fitting the low-frequency part of the power spectrum for HL CMa in its quiescent state by a function of the form $nf^{\alpha}(1 + (f/f_{break})^4)^{\beta}$. 
As a result of the fitting, we obtained the parameters $\alpha \approx -1$ and $\beta \approx 0.4$. 
The variability amplitude for HL CMa in the power spectrum derived from the light curve in the quiescent state is higher than that in the two other observations. 
Therefore, the shape of the model function describing the variability of HL CMa can be estimated without allowance for the atmospheric variability.

As has been noted above, there were no stars brighter than magnitude 13 in the telescope's field of view. 
The absence of bright constant stars makes it difficult to estimate the shape of the atmospheric variability, because a significant contribution to the variability from Poissonian noise causes the uncertainty to increase in the entire frequency range \footnote{The points of the power spectrum have $\chi_2^2$,statistics; the dispersion in the frequency range($f_i$) is equal to the variability power density in this range ($P_{i} = m(f_i)$). 
The uncertainty in estimating the parameters of the atmospheric variability model $n_{atm}, f_{atm}$, is determined by the combined variability power $n_{atm} + P_{pois}$. 
The uncertainty in estimating the atmospheric model parameters increases with Poissonian noise amplitude.}. 
The gray line on the graphs of the power spectra presented in Fig. \,\ref{fig:hl_cma_ps}, marks the uncertainty region of the power spectrum for the differential light curve of the constant stars after the subtraction of the atmospheric model. 
The shape of the power spectrum for HL CMa under this curve cannot be determined accurately due to the uncertainty of the atmospheric noise model.
	
\bigskip

\subsection*{System Geometry}	
	
In the off state, the break in the brightness variability power spectrum is at the frequency$f_{b}=(6.5 \pm 1.0) \times 10^{-3}$ Hz (the upper panel in Fig.\,\ref{fig:hl_cma_ps}). 
If this break corresponds to the Keplerian time at the inner edge of the optically thick disk, then its radius is $R_{in} = \sqrt[3]{(GM_{wd})/(4 \pi^2 \nu^2)} \approx 4 \times 10^{9}$cm,corresponding to seven white dwarf radii. In our calculations, we took the white dwarf mass and radius to be $M_{wd} = M_\odot$ \citep{hutchings81}, and $R_{wd}=6\times 10^8$ cm \citep{gatewood78}. 
At maximum optical light (magnitude 12, two days after the outburst onset), the break in the power spectrum was approximately at the same frequency $f_{b} = (7.0 \pm 2.0) \times 10^{-3}$ Hz (the lower panel in Fig.\,\ref{fig:hl_cma_ps}). 
In this case, it should be noted that the break frequency at the decay stage is slightly shifted toward lower value, $f_{b} = (3.5
\pm_{1.0}^{1.2}) \times 10^{-3}$ Hz,but it remains compatible with the derived values of the break frequency in the bright and off states within the $2 \sigma$ confidence interval.

As has been said above, in the evaporating disk model \citep{meyer94} , the inner edge of the optically thick disk cannot reach the white dwarf surface. 
During an outburst, the inner edge of the optically thick disk must move closer to the white dwarf surface due to the increased accretion rate. 
This must lead to an increase in the frequency at which the system?s luminosity variability is suppressed. 
Such behavior is observed, for example, in the dwarf nova SS Cyg\citep{revnivtsev12, balman12}.
	
In contrast to SS Cyg, the observations of HL CMa show no pronounced change of the variability suppression frequency in the quiescent state compared to the outburst one. The prolonged optical observations of the dwarf nova V1504 Cyg \citep{dobrotka15}, performed by the {\it Kepler} observatory, just as in the case of HL CMa, show no significant changes in the break frequency of the brightness variability power spectrum in its outburst and quiescent states. 
It should be noted that the Kepler observations have several advantages over the observations with ground-based telescopes: the possibility of prolonged observations (more than a day) and the absence of variability related to the properties of the atmosphere.
This leads to a high accuracy of the estimates for the break frequency in the power spectrum.

\section{DISCUSSION}

%====================================================================================
\begin{figure*}
\includegraphics[width=1\textwidth, height=0.28\textheight]{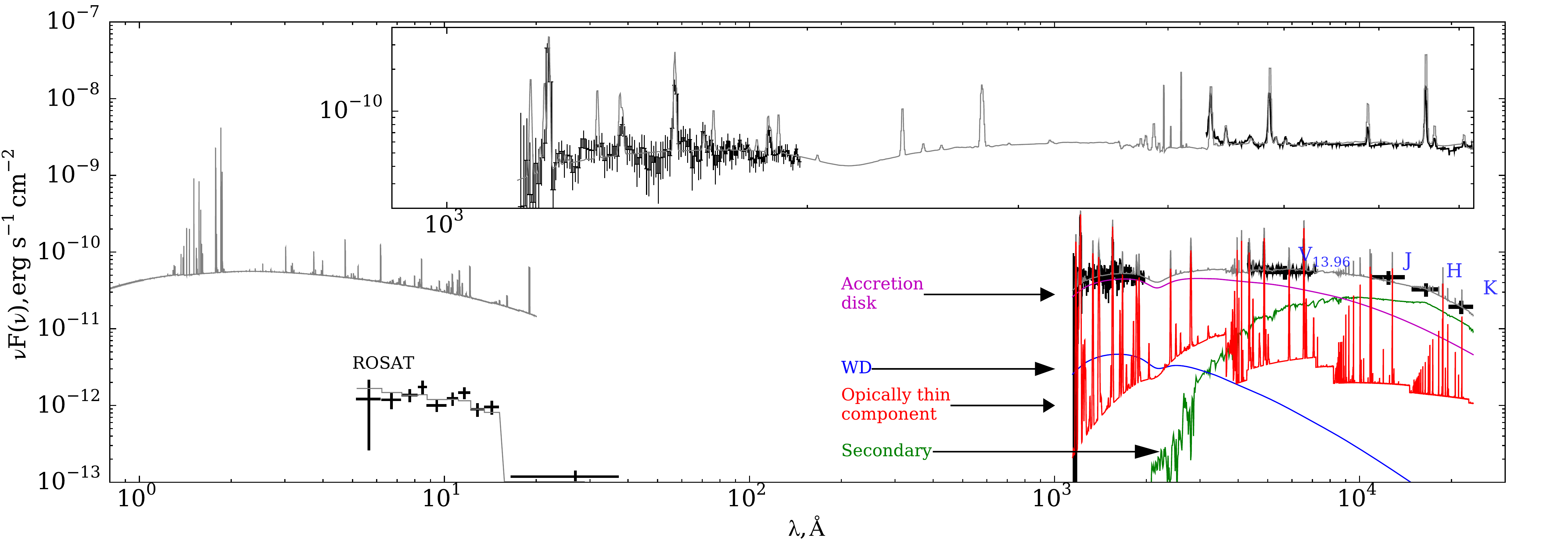}
\includegraphics[width=1\textwidth, height=0.28\textheight]{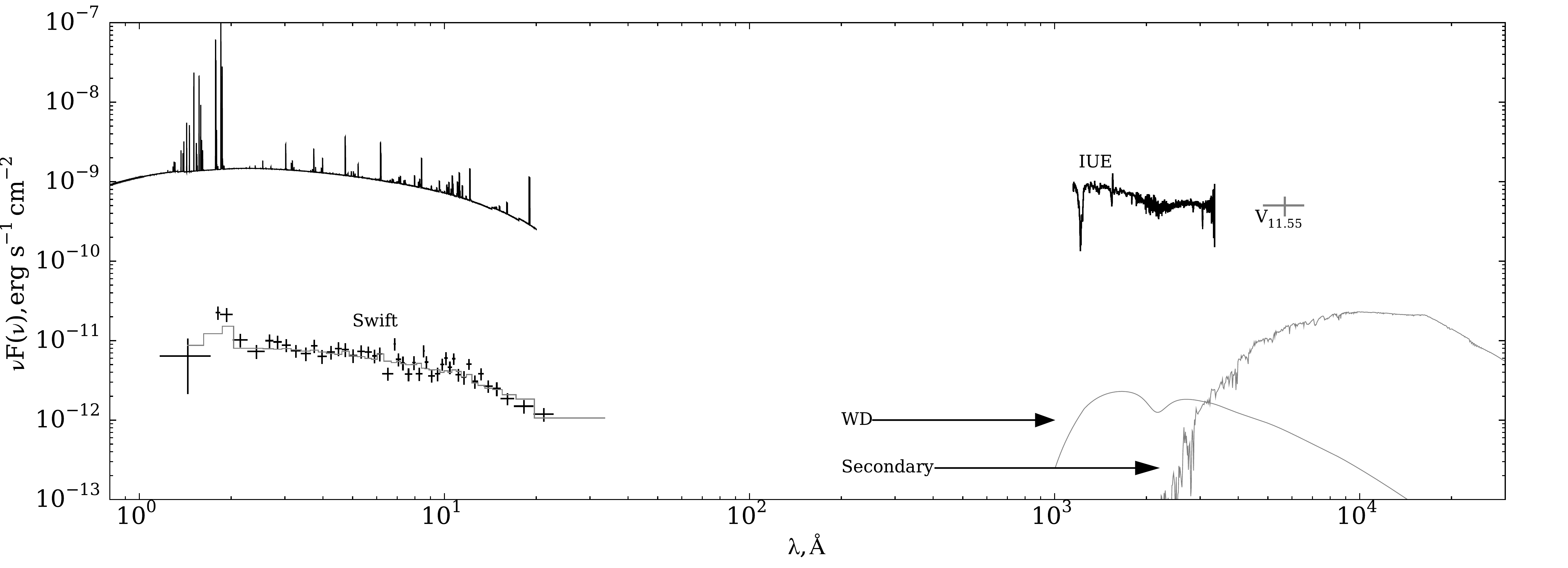}
\captionsetup{singlelinecheck=off}
\caption[]{Upper panel: infrared \citep[][J, H, K;]{szkody87} and optical (V ) photometry, the ultraviolet ({\it IUE}) and X-ray ( {\it ROSAT}) spectra in the quiescent state of HL CMa, the optical spectrum was taken from \cite{sheets05}. 
Lower panel: the ultraviolet ({\it IUE}) and X-ray ({\it Swift}) energy spectra during its outburst. The measurements are indicated by the black curves and crosses. 
The gray curve on the upper panel in the low-energy part of the spectrum represents the system?s model spectrum consisting of several components: a white dwarf, a blackbody disk, an optically thin flow, and a companion star (see the text). The part of the spectrum from 1000 to 7300 \AA is presented in the upper part of the upper panel. 
The model of emission from an optically thin hot plasma with a temperature of 8 keV and a luminosity corresponding to the accretion rate in the outer disk as well as the model corresponding to the observed flux from {\it ROSAT} data (gray curves) are presented in the X-ray part of the spectrum. On the lower panel in the X-ray part of the spectrum, the gray curve represents the model of emission from an optically thin plasma with a temperature of 8 keV and a luminosity corresponding to the accretion rate $\dot{M} = 1.2 \times 10^{-8} M_\odot/$yr$^{-1}$. The blackbody spectrum of the white dwarf and the companion star is presented in the low-energy part of the spectrum. The ultraviolet spectrum of the system in its bright state is determined by the disk and wind radiation in
the system \citep[see ][]{mauche87}.The distance to the system was taken to be 250 pc.
}
\label{fig:wb_spec}		
\end{figure*}
%===========================================================================

A change in the variability properties of dwarf novae during their transitions between different states is expected from a change in the physical properties of the accretion disk. 
However, in this paper we have shown that the characteristic variability power suppression frequency in HL CMa barely changes when it passes from the quiescent state to the outburst one.
The the increase of the variability power suppression frequency predicted in the evaporating disk model was observed in the similar system SS Cyg.
	
The difference in variability evolution between SS Cyg and HL CMa during the transition from the quiescent state to the outburst one cannot be explained only on the basis of differences in the power
spectra, because the mechanism of their formation is not known completely. 
To qualitatively determine the difference between these systems, we analyzed the available spectroscopic observations of HL CMa in a wide frequency range.
	
A detailed analysis of the system was performed by \citet{mauche87} using ultraviolet spectra and optical and near-infrared photometry. They showed the distance to the system to be $D = 250 \pm 50$ pc, and the accretion rate in its bright state to be $\dot{M} \approx 10^{-8} M_{\odot}$. 
The accretion rate was estimated from the shape and normalization of the ultraviolet spectrum, V-band luminosity, and orbital period of the system. Subsequently, periods of prolonged intermediate states were detected in the system, which transfer it to the class of so-called Z Cam systems\citep{mansperger94},characterized by a
high accretion rate.

In the ultraviolet spectral range, 57 spectra were taken by the IUE observatory in the period from 1983 to 1993. 
To construct the averaged ultraviolet spectra in the quiescent state, we used six spectra (taken in period when the V magnitude was greater than 13). The mean V magnitude for these spectra turned
out to be 13.96 (Fig.\,\ref{fig:wb_spec}). 
The averaged spectrum in the outburst state was made from 18 spectra taken in periods when the system?s brightness was higher than magnitude 12 in the V band. 
The mean V magnitude for these spectra is 11.55. 

In the infrared part of the spectrum, we used Kitt Peak J, H, K photometry \citep[see Table 1 in][]{szkody87}. The observations were carried out in February 1984 (11 days after the onset of another outburst, when the apparent V magnitude was about 14).
	
In the optical part of the spectrum, we used data from the AAVSO archive. We also used the spectrum in the wavelength range 4300 - 7000 \AA, taken from \cite{sheets05}.
	
Several X-ray observations of HL CMa were performed:
\begin{itemize}	
\item the {\it PSPCb/ROSAT} observations (February 15, 1991, and April 20, 1991, in the quiescent state, the total exposure time of the two observations is 3231 s);	
\item the {\it XRT/Swift} observations (performed from May 31, 2015, to June 3, 2015, when the system was near its maximum light, the total exposure time is 6400 s).
\end{itemize}
	
The {\it ROSAT} and {\it Swift} data reduction was done
using the heasoft-6.17 software package (https://
heasarc.gsfc.nasa.gov/lhea- soft/). The spectra were
analyzed using the software package {\sc Xspec} \citep{arnaud96}.
	
\bigskip

\subsection*{Spectrum}

The {\it Swift} spectrum in the energy range $0.5-10$ keV is well fit by the model of absorbed emission from a single-temperature optically thin plasma ({\sc wabs*mekal} in the {\sc Xspec}) with the following parameters: $N_{\rm H} = (0.11 \pm 0.02) \times 10^{22}$ cm$^{-2}$, $kT = 8.1 \pm 1.4$ keV. 
The recorded flux from the source is $0.5-2$ keV $F_{0.5-2} \approx  4.5 \times 10^{-12}$ erg s$^{-1}$ cm$^{-2}$, in the $0.5-10$ keV energy band and $F_{0.5-10} \approx 1.8 \times 10^{-11}$ erg sec$^{-1}$ cm$^{-2}$. 
The measured parameters of the spectrum allow the total unabsorbed flux to be estimated in the wide X-ray energy range $0.1 -100$ keV  $F_{tot} \approx 2.74 \times 10^{-11}$ erg sec$^{-1}$ cm$^{-2}$.
	
The X-ray flux in the quiescent state was measured with the PSPC instrument of the {\it ROSAT} observatory operated in the $0.1 - 2.4$ keV. 
Since such an energy band does not allow the plasma temperature to be reliably estimated, the spectrum was fit by a model with only one free parameter, the normalization. 
The plasma temperature was taken to be $kT = 8$keV (it corresponds to that measured in the outburst state); the absorption column density was taken to be$N_{\rm H} = 0.51\times 10^{21}$ cm$^{-2}$, corresponding to the extinction $E(B - V) = 0.1$, from \citep{mauche87}. 
This value agrees well with the present-day extinction maps \citep{lallement14}. 
The measured X-ray flux from HL CMa in the quiescent state was $F_{0.5-2} \approx 1.37 \times 10^{-12}$ erg sec$^{-1}$cm$^{-2}$ in the $0.5 - 2$ keV energy band. 
This allows the total unabsorbed flux in the $0.1 - 100$ keV energy band to be estimated, $F_{tot} \approx 7.1
\times 10^{-12}$ erg sec$^{-1}$ cm$^{-2}$, which is a factor $\approx 4$ lower than that in the bright state.

Based on the total luminosity estimated for the model of a single-temperature optically thin plasma in the {\it ROSAT} observations, $L_x \approx 0.5 \times 10^{32} $erg s$^{-1}$, the accretion rate must be $\dot{M} \approx R_{wd}L_{x}/(GM_{wd}) = 2.5 \times 10^{14}$ g/sec. 
This value is much lower than that expected from the optical luminosity, $\dot{M} \approx 2 \times 10^{16}$ g/sec. 
In the outburst state, the accretion rate in the disk rises to $\dot{M} \approx 6 \times 10^{17}$ g/sec, while the X-ray luminosity increases only by a factor of $\simeq 4$. 
Thus, the total X-ray luminosity of HL CMa in both quiescent and outburst states turns out to be much lower than that expected from the accretion rate in the disk.
		
\citet{revnivtsev14}showed that the total X-ray luminosity of cataclysmic variables corresponds to the optical luminosity of these systems under the assumption of a constant accretion rate in
the entire disk. This conclusion was reached for a
sample of cataclysmic variables with low accretion
rates (less than $10^{16}$ g/sec), in which the energy of accreting matter is released near the white dwarf surface through optically thin hot plasma losses.

However, a sharp decrease in the X-ray luminosity is observed in some systems in the bright state \citep[see, e.g.,][]{wheatley96, wheatley03, collins10, ramsay12}. 
Such a peculiarity of these systems is explained, in particular, by the transition of an optically thin boundary layer to an optically thick one at an accretion rate in it higher than $10^{16}$ g/sec \citep{patterson85, piro04}. 
An example of a system showing such a behavior is SS Cyg. 
In the quiescent state, the total X-ray flux from this system is considerably higher than that in its outburst and corresponds to the presumed energy release through accretion. 
At maximum light during an outburst, the X-ray flux drops significantly, but the energy release grows in the hard ultraviolet and soft X-ray parts of the spectrum. 
The total optical luminosity turns out to be compatible with the energy release in the hard ultraviolet range under the assumption that the accretion rate through the disk is constant and equal to the accretion rate on the white dwarf surface.

Figure \,\ref{fig:wb_spec} presents broadband energy spectra of
HL CMa in its quiescent (upper panel) and outburst
(lower panel) states. The gray curve on the upper
panel (quiescent state) in the low-energy part of the ($\lambda$ > 1000 \AA) indicates the model that is a combination of several individual components (also indicated on the graph by the color lines):
\begin{itemize}
\item A companion star (green curve) with a mass of $0.5 M_\odot$, radius $R_{sec} = 3.88 \times 10^{10}$\,cm
    \citep{knigge06}, temperature $T_{sec} = 3750$\,K \citep{eker15}, radiation heated to a temperature of 6000 K across 10\% of its surface. The stellar spectra were taken from the library of spectra \citep{kurucz93}\footnote{The library can be found at http://www.stsci.edu/hst/observatory/crds/k93models.html}.
\item An accretion disk (purple curve) with the spectrum of a multi-temperature $\alpha$--disk \citep{ss73}. 
The accretion rate in the disk is assumed to be constant up to the radius at which the disk temperature $T < 10^4$K, and equal to $\dot{M}_0 =     2.1 \times 10^{-10} M_\odot/$year. 
Within the radius where the $\alpha$--disk temperature is higher $10^4$K, the temperatures correspond to an $\alpha$--disk with an
accretion rate decreasing as $\dot{M} = \dot{M}_0 (0.6 ((r -
    r_{wd})/(r_4 - r_{wd}))^2 + 0.4)$, where $r_4$ -- is the radius
of the disk with a constant accretion rate at which its temperature reaches $10^4$К. 
The disk is assumed to be geometrically thin; locally, the spectrum at each radius corresponds to the spectrum of a blackbody with an effective temperature $T = \left( 3 G M_{wd} \dot{M}(r)/(8 \pi \sigma r^3) \right)^{1/4}$, where $\sigma$ -- is the Stefan-Boltzmann constant; The disk luminosity was calculated by taking into account the disk luminosity distribution in angle $1/2(1 + 3/2 \cos{\theta})$ \citep{mauche87}.
\item Optically thin plasma emission (red curve) with temperatures $T = 18000$\,K (luminosity $5 \times 10^{32}$ erg sec$^{-1}$), $T = 40000$\,K (luminosity $8 \times 10^{31}$ erg sec$^{-1}$) and $T = 3 \times 10^5$\,K (luminosity $7 \times 10^{31}$ erg sec$^{-1}$). 
The optically thin emission spectrum was obtained with the Cloudy software package \citep{ferland98}. The optically thin components in our model corresponded to an optically thin flow above the disk. To obtain the spectrum of these components, the Cloudy package computed a spherical shell with thickness
$h = 2 \sqrt{(k_b T r_m^3)/(G M_{wd} m_p)}$, where $k_b$ -- is the Boltzmann constant, $T$ -- is the shell temperature, $G$ -- is the gravitational constant, $m_p$ -- is the proton mass, and $r_{m}$ -- is the radius of the disk at which its temperature reached 18000\,K (if the accretion rate remained constant $\dot{M} = \dot{M}_0$). 
The inner radius of the spherical shell was taken to be $10^{10}$ cm much larger than the shell thickness). 
There was a blackbody radiation source with a temperature of 18000\,K and a radiation intensity on the surface of the spherical shell equal to the radiation intensity of an $\alpha$--disk with a temperature of $18000$\,K at the center of the spherical shell. 
The temperature and density in the shell were assumed to be constant, while the density was chosen in such a way that the surface luminosity of the shell was 20\% of the surface luminosity of an $\alpha$--disk with a temperature of 18000\,K. 
The normalization of the spectra in the final model was chosen to describe the emission lines in the observed ultraviolet and optical spectra. 
\item A white dwarf (blue curve) with a blackbody spectrum at a temperature of 25000 K.
\end{itemize}

In the described spectral model, the optically thick disk reaches the white dwarf surface, losing about half of the accretion rate. The presence of inner parts of the optically thick accretion disk turns out to be important for obtaining acceptable slopes of the continuum spectral components in the ultraviolet and optical ranges and for the ratio of the system?s luminosities in these ranges.

In the X-ray part of the spectrum, the gray curve represents the model of a single-temperature plasma with a temperature of 8 keV and a normalization corresponding to the accretion rate $\dot{M}_0$ (under the assumption that an energy $(GM_{wd}\dot{M}_0)/(2 R_{wd})$ is released on the white dwarf surface in the optically thin regime).

Based on the derived spectral peculiarities, we can conclude that the boundary layer of HL CMa in both quiescent and bright states is optically thick. 
This implies that much of the energy release near the white dwarf surface  is emitted as hard ultraviolet and soft X-ray, unobservable due to interstellar extinction. 
This also implies that the accretion rate onto the white dwarf surface in HL CMa in its quiescent state exceeds $10^{16}$ g/sec. 
In view of the high accretion rate, the optically thick accretion disk most likely reaches the white dwarf surface.

In this interpretation, the break frequency in the variability power spectrum for HL CMa does not correspond to the Keplerian time at the inner edge of the optically thick disk that reaches the white dwarf surface. 
Under these assumptions about the shape of the accretion flow, we can propose several possible variability suppression mechanisms: 
\begin{itemize}
\item The optically thick accretion disk reaches the white dwarf surface; the variability is suppressed not on the Keplerian time scales but on the viscous ones at the inner disk edge.
\item The variability suppression is associated with rapid incomplete evaporation of the optically thick disk with a radius of $4 \times 10^{10}$cm (the optically thick disk reaches the white dwarf surface, but much of the accretion occurs in the optically thin flow and has a different variability power spectrum).
\item The variability suppression is associated with the characteristic radiative cooling time in a medium that makes the greatest contribution to the optical luminosity variability, for  example, in the optically thin corona.	
\end{itemize}

\bigskip
This work was financially supported by the Russian Foundation for Basic Research (project nos. 14-02-93965 and 13-02-00741) and the Program of the President of the Russian Federation for Support
of Leading Scientific Schools (project no. NSh-6137.2014.2). We are grateful to V.F. Suleimanov for the fruitful discussion of the mechanisms for the formation of radiation and variability in dwarf novae as well as for the helpful remarks on the paper. We are grateful to P. Wood for the help with the observations.

\bigskip
translation of the article made by V. Astakhov	
\label{lastpage}

\bibliographystyle{unstrnat2} % ГОСТ 7.80
\bibliography{hl_cma_rus2}

\end{document}